\documentclass[amsfonts,twocolumn,showpacs,preprintnumbers,amsmath,amssymb,pra]{revtex4}

\def\weird_offset_after_small{\hskip -1.5mm}   

\def\bbbone{{\bf 1}}
\def\bea{\begin{eqnarray}}
\def\eea{\end{eqnarray}}
\def\ben{\begin{equation}}
\def\een{\end{equation}}
\def\eq#1{Eq.~\ (\ref{#1})}
\def\eqs#1#2{Eqs.~\ (\ref{#1},\ref{#2})}
\def\sss{\scriptscriptstyle\rm}

\def\l{^\lambda}
\def\half{\frac{1}{2}}
\def\bp{{\bf p}}
\def\br{{\bf r}}

\def\ba{{\bf a}}
\def\bq{{\bf q}}

\def\bj{{\bf j}}
\def\xc{_{\sss XC}}
\def\ext{_{\rm ext}}
\def\ee{_{\rm ee}}

\def\bv{{\bf v}}
\def\bA{{\bf A}}
\def\H{_{\sss H}}
\def\hatV{{\hat V}}

\def\hatbp{{\hat \bp}}
\def\hatbr{{\hat \br}}
\def\hatbv{{\hat \bv}}
\def\hatbj{{\hat{\bf\jmath}}}
\def\Vee{{V\ee}}
\def\hatVee{{\hatV\ee}}

\def\extla{_{{\rm ext}_,\lambda}}
\def\subl{{_{1/\lambda}}}
\def\subll{\subl}

\def\unif{^{\rm unif}}
\def\Phio{{\Phi_0}}
\def\Psio{{\Psi_0}}
\def\Phiol{{\Phi_{0,1/\lambda}}}
\def\Psiol{{\Psi_{0,1/\lambda}}}
\def\tddep{[\bj;\Psio,\Phio]}  
\def\tddepl{[\bj\subll;\Psiol,\Phiol]}  

\def\fmxc{\overleftrightarrow{f\xc}}
\def\fmlxc{\overleftrightarrow{f\l\xc}}
\def\chim{\overleftrightarrow\chi}
\def\chiml{\overleftrightarrow{\chi\l}}
\def\chism{\overleftrightarrow{\chi_s}}
\def\Tm{\overleftrightarrow T}
\newcommand{\trace}[1]{{\rm Tr}\left(#1\right)} 
\def\ddt{{\frac{\partial}{\partial t}}}
\def\ddtp{{\frac{\partial}{\partial t'}}}
\def\PRL{Phys. Rev. Lett.\ }
\def\dst{(\br,\br',t-t')}   
\def\dstl{(\lambda\br,\lambda\br',\lambda^2(t-t'))} 
\def\vnabla{\vec\nabla}

\begin{document}

\title{Coordinate scaling in time-dependent current density functional theory}

\author{Maxime Dion and Kieron Burke}
\affiliation{Department of Chemistry and Chemical Biology, 
Rutgers University, 610 Taylor Rd., Piscataway, NJ 08854, USA}

\date{\today}

\begin{abstract}
The coupling constant dependence is derived
in time-dependent {\em current} density functional theory.
The scaling relation can be used to check approximate functionals
and in conjunction with the adiabatic connection formula 
to obtain the ground-state energy from the exchange-correlation kernel.
The result for the uniform gas using the Vignale-Kohn approximation
is deduced.
\end{abstract}


\maketitle


Time-dependent functional density functional theory (TDDFT) \cite{RG84}
is developing rapidly as a tool for predicting electronic response
to laser fields, both weak and strong \cite{WGB05}.
For weak fields, linear
response applies, and perhaps the most popular present application
of TDDFT is in calculating optical response of molecules, including
transition frequencies \cite{FB04}.  For strong fields, TDDFT allows prediction
of many properties in response to intense laser pulses, such as high
harmonic generation \cite{MG04}.  It seems likely that TDDFT will play a key
role in the emerging field of electron quantum control \cite{WGB05}.

The time-dependent scheme can also be used for strictly ground-state 
properties by using the
adiabatic connection formula to compute the static exchange-correlation
energy \cite{LP75,GL76,LP77}.  
Although using TDDFT to calculate ground-state properties
might seem an unwarranted complication, the approximate 
exchange-correlation energy
functional within this scheme has many useful properties.  For example,
it correctly describes the static correlation for bond dissociation
\cite{FNGB05} and can be used to calculate accurately
van der Waals dispersion energies \cite{OGSB97,KMM98,MJS03,DRSL04}.

To use TDDFT in the adiabatic connection formula, one must generalize
the response functions to arbitrary coupling constant $\lambda$. 
In DFT, this has a very precise meaning, as the density is held fixed
while the Coulomb repulsion between 
electrons is multiplied by $\lambda$ \cite{LP75,GL76}.
Coordinate-scaling is used to derive the $\lambda$-dependence of quantities
in DFT \cite{LP85}, 
or {\em current} DFT \cite{EG96}.  
It leads to many useful results, such as the
virial theorem for the exchange-correlation energy \cite{LP85}, 
and exact conditions on that energy \cite{L91}.
It can be used to check that approximate functionals have the right scaling
behavior \cite{L95}.

Time-dependent {\em current} density functional theory (TDCDFT)
is a more general scheme where the current density is the basic variational
parameter instead of the density and it can include arbitrary magnetic fields.
Unlike TDDFT, TDCDFT can be approximated with local or 
semi-local functionals without any conceptual difficulty \cite{VK96}
and is now being used to calculate excitations of quantum wells \cite{UV98,UV01}, 
atoms \cite{UB04},
molecules \cite{FBLB02,FBLB03,FBb04,FBc04} and
single molecule transport \cite{BCG05,BKE05}.
Five years ago, the connection between coordinate-scaling and the
coupling constant was derived for TDDFT \cite{HPB99}.
The present paper generates analogous results within the more
general framework of TDCDFT.

TDCDFT \cite{GD88,V04} starts from the Schr\"odinger
equation for $N$ electrons in a vector potential $\ba\ext$:
\ben
\left\{ \half \sum_{i=1}^N \left(\hatbp_i+\ba\ext(\hatbr_i,t) \right)^2 + 
\hatV\ext + 
\hatVee \right\} \Psi
= i\ddt \Psi
\label{schro_vector}
\een
where $\hatV\ext$ denotes the one-body potential,
\mbox{$\hatVee=\frac{1}{2}\sum_{i,j=1}^{N} |\hatbr_i-\hatbr_j|^{-1}$}
and $\hatbp_i=-i \hat\nabla_i$.  
We use atomic units throughout ($e^2=m_e=\hbar=1$), and
there is an implicit speed of light constant, $c$,
included in the vector potential, i.e., ${\bf B}=c \nabla \times \bA$, where
${\bf B}$ is the usual magnetic field.
The physical results from the Schr\"odinger
equation above are invariant under the gauge transformation
\begin{eqnarray}
\bar v\ext(\br,t) & = & v\ext(\br,t) - 
\frac{\partial \Lambda(\br,t)}{\partial t} 
\label{gauge1}
\\
\bar \ba\ext(\br,t) & = & \ba\ext(\br,t) +
{\bf\nabla} \Lambda(\br,t),
\label{gauge2}
\end{eqnarray}
where $\Lambda$ is an arbitrary function.
The gauge freedom can be used, for example, to remove of the scalar
potential by choosing
$\partial\Lambda(\br, t)/\partial t = v\ext(\br,t)$.

As the density is the conjugate variable to the scalar potential 
$v\ext(\br,t)$,
the conjugate variable to the vector potential is the current density
\ben
\hatbj(\br,t)=\half \sum_{i=1}^N\left\{ 
\hatbv_i(t) \delta(\br-\br_i) + \delta(\br-\br_i) \hatbv_i(t) \right\},
\een
where $\hatbv_i(t)=\hatbp_i+\ba\ext(\hatbr_i,t)$.
The basic theorem of TDCDFT \cite{GD88,V04} 
states that, for a given initial wavefunction,
a given $\bj(\br,t)$ is generated by at most one $\ba\ext(\br,t)$, 
up to a gauge transformation.
The density and the current are related through the continuity equation
\ben
\frac{dn(\br,t)}{dt} + {\bf\nabla} \cdot \bj(\br,t) = 0
\label{continuity}
\een

To derive the coupling constant dependence, we first
transform the coordinates to 
\mbox{$(\br_i,t)=(\lambda\br_i',\lambda^2 t')$}
in \eq{schro_vector}:
\begin{small}
\ben
\label{schro_scaled1}
\left\{ \half 
\sum_{i=1}^N \left( \frac{\hatbp_i'}{\lambda} +
\ba\ext(\lambda\hatbr_i',\lambda^2 t') \right)^2 + 
\hat V'\extla + 
\frac{\hatVee'}{\lambda} \right\} \Psi_\lambda' = 
\frac{i}{\lambda^2} \ddtp \Psi_\lambda'.
\een
\end{small}
\weird_offset_after_small
where the prime means that the quantity is evaluated at $(\br',t')$, 
\mbox{$\hat V'_{{\rm ext},\lambda}=\sum_{i=1}^N 
v\ext(\lambda \hatbr_i',\lambda^2 t')$},
and the scaled normalized wavefunction is
\mbox{$\Psi_\lambda'= 
\lambda^{3N/2} \, \Psi(\lambda \br'_1... \lambda \br'_N, \lambda^2 t')$}.
Consistent with Ref.~\ \cite{HPB99},
we define the scaled density by
\mbox{$n_\lambda(\br,t)=\lambda^3\, n(\lambda\br,\lambda^2 t)$}.
Now we also define the scaled current density
\ben
\bj_\lambda(\br,t)=\lambda^{4}\, \bj(\lambda\br,\lambda^2 t).
\een
Continuity (\eq{continuity}) remains satisfied for all $\lambda$.
Multiplying \eq{schro_scaled1} by $\lambda^2$ and omitting the primes, 
\begin{small}
\ben
\left\{ \half 
\sum_{i=1}^N 
\left(\hatbp_i+\lambda \, \ba\ext(\lambda\hatbr_i,\lambda^2 t)   \right)^2 + 
\lambda^2 \hatV\extla + 
\lambda \hatVee \right\} \Psi_\lambda = 
i\ddt \Psi_\lambda.
\label{schro_scaled}
\een
\end{small}
\weird_offset_after_small
We define $\ba\l\ext[\bj,\Psi_0]$ as the vector potential 
for a system with modified
coupling constant $\lambda$, which gives rise to current $\bj$ starting from
wavefunction $\Psi_0$.
Thus we identify
\ben
\ba\ext\l[\bj_\lambda,\Psi_{0,\lambda}](\br,t) = 
\lambda \, \ba\ext[\bj,\Psi_0](\lambda \br,\lambda^2 t).
\label{aext_scaling}
\een
Although the $\lambda$-dependence of the external potentials 
in \eq{schro_scaled} is generally complicated, by virtue
of the one-to-one correspondence between current and potentials
\cite{GD88,V04}, the vector potential
appearing in \eq{schro_scaled} is that {\em unique} potential 
producing current density $\bj(\br,t)$ from the initial wavefunction $\Psi_0$,
with electron-electron interaction $\lambda \hatV\ee$.

Next we apply the same argument to the Kohn-Sham system, where 
the electrons are non-interacting \mbox{($\hatV\ee=0$)} and
$\ba\ext(\br, t)$ is replaced by an effective vector 
potential, $\ba_s(\br, t)$, defined to 
reproduce the same current as the interacting system.  
Since our previous argument does not depend on the interaction,
$\ba_s^\lambda(\br, t)$ also satisfies \eq{aext_scaling}.
And the Hartree vector potential,
\mbox{$\ba\H(\br, t)=\vec\nabla \int^{t} dt' 
\int d\br'\; e^2\; n(\br',t')/|\br-\br'|$},
satisfies the same scaling.
From the definition of the exchange-correlation potential, 
\mbox{$\ba_s=\ba\ext+\ba\H+\ba\xc$}, 
we see that it must obey the same scaling as the other vector potentials:
\ben
\ba\xc\l\tddep(\br, t) = \lambda \; \ba\xc\tddepl(\lambda\br,\lambda^2 t),
\label{axc_scaling}
\een
where there is also a functional dependence on the initial Kohn-Sham 
wavefunction $\Phi_0$, from $\ba_s(\br, t)$.
This is the central result of this work.

When the vector potential is irrotational, i.e.\ can be gauge-transformed
to a scalar potential, the TDDFT $\lambda$-dependence of Ref.~\ \cite{HPB99}
can be derived from these more general results.  
From the gauge transformation, \eqs{gauge1}{gauge2}, one can see that an 
irrotational vector potential is transformed to a scalar potential through 
$\partial \ba/\partial t= \nabla v$.  Inserting \eq{axc_scaling},
we find
\ben
\nabla v^\lambda\xc(\br,t)=\lambda \,
\partial \ba\xc(\lambda \br, \lambda^2 t)/\partial t
= \lambda^3 \, \nabla_\lambda v\xc(\lambda \br,\lambda^2 t),
\een
where $\nabla_\lambda= \partial/\partial(\lambda \br)$.  
Requiring the potential to vanish far from the system we recover
$v\xc^\lambda(\br,t)=\lambda^2 v\xc(\lambda \br, \lambda^2 t)$
from Ref.~\ \cite{HPB99}.  This relation can also be derived directly
from \eq{schro_scaled} by the same arguments used for the vector potential.

While \eq{aext_scaling} represents the most general form,
applicable to all TDCDFT applications, we next look at the special case of
the linear response of an electronic system.
The susceptibility, $\chi$, is usually defined by
\mbox{$\delta n(\br,t)=\int d\br' dt' \; 
\chi(\br,t;\br',t') \; \delta v(\br',t')$},
where $\delta n$ is a small change in
density due to a small perturbation in the potential, $\delta v$.
We sometimes represent the previous equation as
$\delta n= \chi * \delta v$.  
Since we now have a vector potential, we can generalize the linear response to
$\delta \bj = \chim * \delta \ba$.
We restrict ourselves to applying (time-dependent)
perturbations on systems for which the external potentials are static.
The response can then be considered as a functional of the ground-state density
only, not the current.

The scaling relation for the linear response exchange-correlation kernel 
in TDDFT is given in Ref.~\ \cite{LGP00}.
In TDCDFT the tensor analog is defined as $\fmxc=\delta \ba\xc/\delta \bj$ and
we can find the scaling relation with the functional differentiation
\begin{small}
\bea
\lefteqn{\ba\xc\l[n+\delta n](\br,t)-\ba\xc\l[n](\br,t) =} 
\nonumber
\\ & &
\lambda (\ba\xc[n\subl+\delta n\subl](\lambda\br,\lambda^2 t)-
\ba\xc[n\subl](\lambda\br,\lambda^2 t) ) =  
\nonumber
\\ & &
\lambda \int d\br' dt' \
\fmxc[n\subl](\lambda\br,\br',\lambda^2 t-t') \ \delta \bj\subl(\br',t') =
\nonumber
\\ & &
\lambda \int (\lambda^3 d\bar\br) (\lambda^2 d\bar t) \
\fmxc[n\subl](\lambda\br,\lambda\bar\br, \lambda^2 (t-\bar t)) \
\frac{\delta \bj(\bar\br,\bar t)}{\lambda^4} 
\nonumber
\eea
\end{small}
\eq{axc_scaling} implies
\begin{small}
\begin{eqnarray}
\fmlxc[n_0]\dst & = & 
\lambda^2\; \fmxc[n_{0,1/\lambda}]\dstl,
\label{fxc_scaling}
\end{eqnarray}
\end{small}
\weird_offset_after_small
or, in frequency space,
\begin{small}
\begin{eqnarray}
\fmlxc[n_0](\br,\br',\omega) & = & 
\; \fmxc[n_{0,1/\lambda}](\lambda\br,\lambda\br',\omega/\lambda^2).
\label{fxc_scaling_omega}
\end{eqnarray}
\end{small}
\weird_offset_after_small
These results are needed to implement the TDCDFT version of the adiabatic
connection formula as shown below.

In the special case of a uniform electron gas
\begin{small}
\begin{eqnarray}
\fmlxc[n_0](\bq,\omega) & = & 
\frac{1}{\lambda^3}\; \fmxc[n_{0,1/\lambda}]\left(
\frac{\bq}{\lambda},\frac{\omega}{\lambda^2} \right).
\label{fxc_scaling_homo}
\end{eqnarray}
\end{small}
\weird_offset_after_small
The above relation implies that, for a uniform gas,
knowing the exchange-correlation kernel as a functional
of the density is the same as knowing the coupling constant dependence;
this was used for the equivalent TDDFT case \cite{APDT03,LGP00}.

There have been various approximations proposed for 
$f\xc$ \cite{GK85,GK86e,PGG96,RA94}
and $\fmxc$ \cite{NCT98,QV02} since they are such important quantities.
The main TDCDFT approximate functional currently in use 
is the Vignale-Kohn (VK) functional \cite{VK96}.
This is the gradient expansion in the current density, and uses
as input the $q \rightarrow 0$ limit of both the longitudinal
exchange-correlation kernel, $f\xc^{\rm L}(\omega)$ (which is precisely the
scalar $f\xc(\omega)$ of TDDFT), and the transverse kernel,
$f\xc^{\rm T}(\omega)$ of the uniform gas.
We have checked that the
VK functional respects the above scaling relation, 
\eq{fxc_scaling_omega}, provided that $f\xc^{\rm \{L,T\}}(\omega)$
used in constructing the functional also respect the appropriate scaling.
The most recent approximation for these kernel components 
is that of Qian and Vignale \cite{QV02}.
We verified that it satisfies \eq{fxc_scaling_homo}, assuming the 
Landau parameters are invariant under simultaneous scaling of the density
and the coupling constant.

Just as for the exchange-correlation potential,
the scaling relation for the exchange-correlation can be derived from TDCDFT.
When the vector potential is irrotational,
the scaling relation of $\fmxc$ reduces to that of the scalar kernel
$f\xc$ \cite{LGP00}, via
\bea
\vnabla \vnabla' f\xc^\lambda[n_0](\br,\br',\omega)  & = &
\omega^2 \fmlxc[n_0](\br,\br',\omega)  \nonumber \\
& = &
\omega^2 \fmxc[n_{0,1/\lambda}](\lambda\br,\lambda\br',\omega/\lambda^2)
\nonumber \\
& = &
\lambda^4 \; \vnabla_\lambda \vnabla'_\lambda
f\xc[n_{0,1/\lambda}](\lambda\br,\lambda\br',\omega/\lambda^2),
\label{fxcrel}
\eea
Then, since $f\xc \rightarrow 0$ as $\br \rightarrow \infty$ for any finite
system, integration implies
than the scaling of the current kernel reduces to the scaling of the scalar
kernel, 
\mbox{$f\xc^\lambda[n_0](\br,\br',\omega)=
\lambda^2 f\xc[n_{0,1/\lambda}](\lambda\br,\lambda\br',\omega/\lambda^2)$}, 
as in Ref.~\ \cite{LGP00}.


Similarly to the adiabatic connection formula used in ground-state DFT 
\cite{LP75,GL76,LP77,FV05},
which relates the exchange-correlation energy to the susceptibility, 
we introduce the adiabatic connection for the ground-state of 
a system with a static scalar potential using {\em current}
DFT susceptibility
\begin{small}
\ben
{\rm E}\xc  =  -\half\int_0^1 d\lambda
\int_{-\infty}^{\infty}\frac{d\omega}{2\pi i} \
\trace{\left[\chiml - n_0(\br) \, \bbbone \right] * \Tm}
\label{acf}
\een
\end{small}
\weird_offset_after_small
the trace is 
\mbox{$\trace{\overleftrightarrow{a}}=\int d\br \sum_{i} \, a_{ii}(\br,\br)$} and
\mbox{$\Tm=-\vnabla \Vee \vnabla/\omega^2$}.
The symbol $\bbbone$ stands for $\delta(\br-\br') \ \delta_{ij}$.
The tensor susceptibility is related
to the exchange-correlation kernel through \cite{VK98}
\begin{small}
\ben
\chiml  =  \chism + 
\chism \left( \lambda\Tm + \fmlxc \right) \chiml,
\een
\end{small}
\weird_offset_after_small
where $\chism$ is the tensor susceptibility for the Kohn-Sham system
\begin{small}
\bea
\chi_{s,ij}(\br,\br',\omega) & = & n_0(\br) \; \bbbone +
\nonumber \\  & &
\sum_{\alpha,\beta} (f_\alpha-f_\beta) \frac{
\phi^*_\alpha(\br) \vnabla_i \phi_\beta(\br) \phi^*_\beta(\br')
 \vnabla'_j \phi_\alpha(\br')}
{\omega-(\epsilon_\beta-\epsilon_\alpha)+i\eta}
\eea
\end{small}
\weird_offset_after_small
where $f$ is the occupation number, i.e. $1$ for an occupied state, 
$0$ for an unoccupied one, and $\eta$ is infinitesimal.  The Kohn-Sham
wavefunctions and energies are denoted by $\phi$ and $\epsilon$.

In the special case of a homogeneous gas, 
the longitudinal and transverse responses decouple, and reordering
the terms within the trace of \eq{acf} shows that only the longitudinal
components contribute to $E\xc$, i.e., it reduces to the usual scalar
case.  Lein et al \cite{LGP00} tested a variety of approximations to
the scalar $f\xc$ for the uniform gas, to see how well they reproduced the
known correlation energy.  To perform the same test for the VK functional,
we first note that, although VK is a gradient expansion in the current,
yielding terms of order $q^2$, these terms are actually zero-order in $q$
when transformed back to the equivalent scalar kernel 
via \eq{fxcrel}.  So we find
that VK, inserted in the current adiabatic connection formula, reduces to 
inserting $f\xc^L(\omega) = f\xc\unif(q\to 0,\omega)$ in the usual scalar
adiabatic connection formula.  This approximation was already tested by
Lein et al, and
is labelled `local RA' in their work.  
(Although they used a different
parametrization \cite{RA94} from QV \cite{QV02}, the results are unlikely to
depend strongly on such details.)
They found about a factor of 2 reduction
in error relative to the adiabatic local density approximation (ALDA).  
We have thus demonstrated that, for the special
case of the uniform gas, the VK approximation, inserted in the current
adiabatic connection formula, improves over ALDA.

Carrying out a calculation of \eq{acf} on molecules or solids
is much more computationally demanding than the usual ground-state
calculations with approximate exchange-correlation
energy functionals, but is probably
not much more expensive than the scalar case.
Such calculations are presently being performed \cite{FG02,FV05}
because the use of the adiabatic connection formula correctly
describes the dissociation of molecules \cite{FNGB05}
and dispersion energies \cite{OGSB97,KMM98,MJS03,DRSL04}.
The exchange-correlation kernel of TDCDFT being better suited to
local or semi-local approximations than the pure density theory \cite{VK96},
we would expect that it would supersede TDDFT when used within the 
adiabatic connection formula.


To summarize, we have used coordinate scaling to derive the coupling-constant
dependence of the exchange-correlation potential in TDCDFT. 
We have derived the adiabatic connection formula for TDCDFT, and shown
how the VK approximation performs for a uniform gas.  We have also given
explicit formulas relating both potentials and kernels in TDCDFT to their
couterparts in TDDFT.
Given both the 
recent use of TDDFT exchange-correlation kernels in the adiabatic
connection formula, for calculating bond dissociation curves, and the 
use and tests of TDCDFT for excitations in which TDDFT has shown limitations,
it is clear that an important application of this work is likely to be 
realized in the near future.

Work supported by NSF grant CHE-0355405.


\end{document}